\documentclass[12pt,preprint]{aastex}

\usepackage{natbib}

\newcommand{\nc}{\newcommand}

\nc{\mpc}{\rm {h^{-1}Mpc }}
\nc{\etal}{{\it et al.\ }}
\nc{\xiav}{\bar{\xi}}
\nc{\omav}{\bar{\omega}}
\nc{\bn}{\bar{N}}
\nc{\tP}{\tilde{P}}
\nc{\tF}{\tilde{F}}
\nc{\avg}[1]{\langle{#1}\rangle}
\nc{\abs}[1]{\mid{#1}\mid}
\nc{\T}[1]{\langle{#1}\rangle_C}

\nc{\eg}{{\it e.g.,\ }}
\nc{\ie}{{\it i.e.\ }}

\nc{\be}[1]{\begin{equation}\mbox{$\label{#1}$}}
\nc{\bea}[1]{\begin{eqnarray} \mbox{$\label{#1}$}}
\nc{\Section}[2]{\section{#2}\label{#1}}
\nc{\Bibitem}[1]{\bibitem{#1}}
\nc{\Label}[1]{\label{#1}}

\nc{\vev}[1]{\langle #1 \rangle}

\nc{\eea}{\end{eqnarray}}
\nc{\ee}{\end{equation}}
\nc{\eeq}{\end{equation}}

\def\lcdm{$\Lambda$CDM~}
\def\ltsima{$\; \buildrel < \over \sim \;$}
\def\gtsima{$\; \buildrel > \over \sim \;$}
\def\simlt{\lower.5ex\hbox{\ltsima}}
\def\simgt{\lower.5ex\hbox{\gtsima}}
\nc\map{{\sl WMAP\ }}

\begin{document}
\title{The Angular Power Spectrum of the First-Year WMAP Data Reanalysed}

\author{Pablo Fosalba\altaffilmark{1}, Istv\'an Szapudi\altaffilmark{1}}

\altaffiltext{1}{Institute for Astronomy, University of Hawaii,
2680 Woodlawn Dr, Honolulu, HI 96822, USA}


\begin{abstract}

We measure the angular power spectrum of the WMAP first-year 
temperature anisotropy maps. We use SpICE
(Spatially Inhomogeneous Correlation Estimator) to estimate $C_\ell$'s 
for multipoles $\ell=2-900$
from all possible cross-correlation channels.
Except for the map-making stage, our measurements provide an 
independent analysis of that by \cite{HinshawEtal2003a}.
Despite the different methods used, there is virtually no difference
between the two measurements for $\ell \simlt 700$ ; 
the highest $\ell$'s are still compatible within $1-\sigma$ errors.
We use a novel {\sl intra-bin variance} method to constrain $C_\ell$ errors in a
model independent way. Simulations show that our implementation of the
technique is unbiased within 1\% for $\ell \simgt 100$.
When applied to WMAP data, the intra-bin variance estimator
yields diagonal errors $\sim 10\%$ larger than those reported by the WMAP team
for $100 < \ell < 450$.
This translates into a 2.4 $\sigma$ detection of systematics 
since no difference is expected between the SpICE and the WMAP team estimator window functions in this multipole range. 
With our measurement of the $C_{\ell}$'s and errors, we get $\chi^2/d.o.f. = 1.042$ for a 
best-fit \lcdm model, which has a 14 \% probability, whereas the WMAP team
\citep{SpergelEtal2003} obtained  $\chi^2/d.o.f. = 1.066$, which has a 5 \% probability.
We assess the impact of our results on cosmological parameters using
Markov Chain Monte Carlo simulations. 
From WMAP data alone, assuming spatially flat power law \lcdm models, 
we obtain the reionization optical depth 
$\tau = 0.145 \pm 0.067$, spectral index $n_s = 0.99 \pm 0.04$, Hubble constant 
$h = 0.67 \pm 0.05$, baryon density $\Omega_b h^2 = 0.0218 \pm 0.0014$, cold dark 
matter density $\Omega_{cdm} h^2 = 0.122 \pm 0.018$, and $\sigma_8 = 0.92 \pm 0.12$,
consistent with a reionization redshift $z_{re} = 16 \pm 5$ (68 $\%$ CL).

\end{abstract}

\keywords{cosmic microwave background --- cosmology: theory --- methods:
statistical}
\section{Introduction}

The {\sl Wilkinson Microwave Anisotropy Probe} satellite (WMAP) 
has provided the clearest view of the
primordial universe to date. Its unprecedented high sensitivity and 
spatial resolution resulted 
in a unique set of cosmic microwave background 
(CMB) radiation maps with close to 
full sky coverage and uniformly high quality.
As a result, fundamental cosmological parameters can
be constrained to the highest precision ever.
Thorough analysis of this dataset \citep{BennettEtal2003a}
yielded a cosmic variance limited measurement of the 
angular power spectrum, $C_\ell$'s, 
of the CMB temperature anisotropy 
for multipoles $\ell \simlt 350$ (\cite{HinshawEtal2003a}; hereafter H03). This
confirmed and improved measurements from previous experiments
(\eg \cite{MillerEtal1999,deBernardisEtal2000,HananyEtal2000,Halverson2002,MasonEtal2003,ScottEtal2003,BenoitEtal2003}).   
The acoustic peak structure revealed by the WMAP temperature and polarization
power spectra provided strong
observational support to inflation and constrained 
viable cosmological scenarios to the domain of flat 
\lcdm models and its close variants.

Considering the importance of these results, our principal aim is to
estimate the angular power spectrum in a completely independent way
in the full range of multipoles probed by WMAP, $2 \le \ell \le 900$, 
and systematically compare results to H03.
Our $C_{\ell}$ estimation pipeline is based on
SpICE \footnote{\rm http://www.ifa.hawaii.edu/cosmowave/}  \citep[Spatially Inhomogeneous Correlator Estimator;][]{SzapudiEtal2001a,SzapudiEtal2001b}, 
a quadratic estimator based on correlation functions.
SpICE performs edge corrections and heuristic minimum variance 
weighting in pixel 
\footnote{The harmonic space alternative using pseudo $C_{\ell}$'s 
is MASTER \citep{HivonEtal2002}.} 
space to produce nearly optimal results.
Our fast HEALPix \footnote{\rm http://www.eso.org/science/healpix/} 
implementation of SpICE  scales as ${\rm {\cal O} (N^{3/2})}$ 
(${\rm N}$ is number of pixels).

\section{Power Spectrum Estimation}
\label{sec:ps}

Our estimation methodology closely follows that of H03,
but adapted to our technique: 

{\em Step 1:} We use
the {\sl foreground cleaned intensity maps} 
for the 3 highest frequency bands Q, V \& W downloaded from the
LAMBDA website \footnote{\rm http://lambda.gsfc.nasa.gov/}. 
Strong diffuse Galactic emission and resolved point sources are masked out
using the Kp0 and Kp2 masks, that  leaves $76.8\%$ and $85.0\%$ 
of the sky useful for cosmological analyses, respectively. 
Monopole $\ell=0$ and dipole $\ell=1$ terms are also removed from 
non-masked pixels.
 
{\em Step 2:} Power-spectrum estimation is  performed via SpICE: 
we compute the cross-correlations from 28 different pairs of channels 
constructed from the 8 ``differencing assemblies'' (DAs) 
Q1 through W4. Noise correlation among different
channels is negligible, therefore our cross-power estimator is unbiased 
with respect to the noise (see \eg H03).
Like H03,
we implement an heuristic $\ell$-dependent 
pixel noise weighting scheme that minimizes errors:
we use flat weights (mask weight only) for
$\ell < 200$, inverse pixel noise variance for $\ell > 450$, 
and a transitional inverse rms noise weight 
in the intermediate range $200 < \ell < 450$.

{\em Step 3:} 
A model for the power spectrum for unresolved extragalactic radio sources
is subtracted from the cross-power
spectrum of each channel. We implement the model given in \S3.1 of 
H03.

{\em Step 4:} 
$C_\ell$'s from different 
channels are optimally combined using an inverse noise 
weighting,  with DA sensitivities as described in the LAMBDA website. 
All channels are included, except for those in Q-band that are only used
in the intermediate $\ell$-range. This helps minimizing galactic contamination
at low $\ell$ and the window function cut-off at the highest multipoles.

{\em Step 5:} 
Our quadratic estimator is defined in pixel space, 
where mask effects 
can be easily corrected for \citep[cf.][]{SzapudiEtal2001a}.
The two point correlation function  is then transformed 
into harmonic space via Gauss-Legendre quadrature 
to obtain the $C_{\ell}$'s deconvolved from
the window function of the experiment.
Symmetrized non-Gaussian beam transfer profiles 
\citep{PageEtal2003}
and pixel window functions are corrected for in $\ell$-space.

\section{Principal Results}
\label{sec:res}

\begin{figure}[htb]
\figurenum{1}
\epsscale{1.}
\plotone{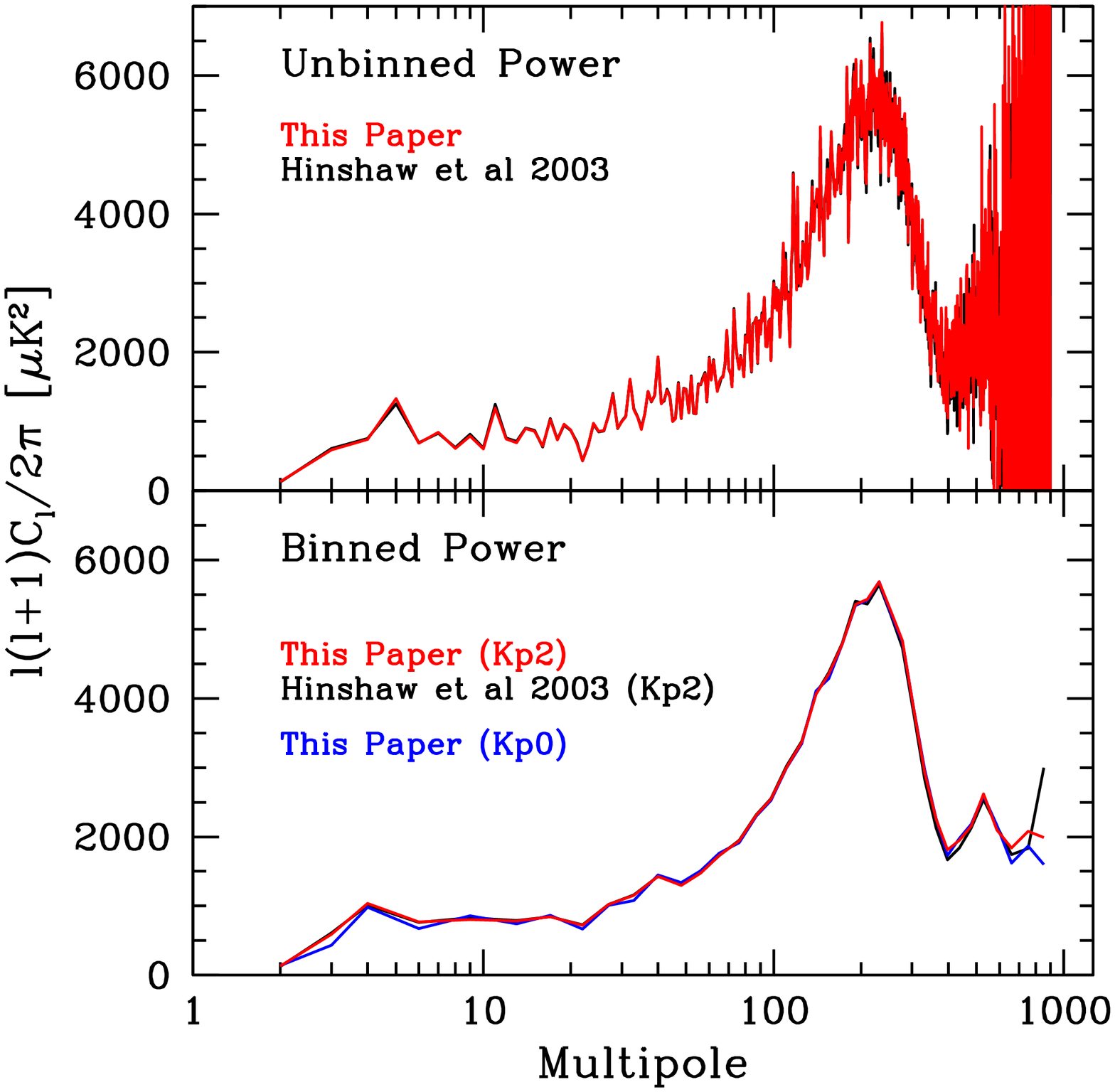}
\caption{WMAP angular power spectrum of the 1st-year 
temperature anisotropy data. 
{Upper panel:} Measurement for individual multipoles (red line) 
using the Kp2 sky cut.
Results from by H03 are also shown (black line,
nearly indistinguishable from red). The agreement is
excellent for most $\ell$'s.
{Lower panel:} Binned power spectrum for two different sky cuts, 
Kp2 (red) \& Kp0 (blue).
We find excellent agreement with H03 
(black line) for all multipoles 
$\ell \simlt 700$ and a slightly different amplitude for the highest $\ell$'s. 
\label{fig:cls}}
\end{figure} 

Figure~\ref{fig:cls} shows the angular power spectrum of WMAP, 
$\Delta T^2_{\ell} \equiv \ell(\ell+1)C_{\ell}/2 \pi$, in $\mu$K$^2$ units, 
measured with SpICE.
Upper panel shows the power spectrum 
for individual multipoles, using Kp2 sky cut.
Our measurement (red line) 
is in excellent agreement with H03
(black line), multipole by multipole. 
In particular, for the quadrupole and octopole 
we find  $\Delta T^2_{2} \sim 135 \mu$K$^2$ and 
$\Delta T^2_{3} \sim 591 \mu$K$^2$, respectively
(H03 get $\sim 123\mu$K$^2$ and $\sim 612 \mu$K$^2$).
For the highest ${\ell}$'s we find slightly different amplitudes 
than H03,
but consistent at the 1-$\sigma$ level.

For the most part, we observe no systematic dependence of the measured 
$C_{\ell}$'s on the sky cut (see difference between red and blue 
lines in bottom panel of Figure~\ref{fig:cls}).
However, 
using Kp0 instead of Kp2 yields
a $15\%$ lower amplitude of the octopole $\ell=3$ and a $15-20\%$
smaller amplitudes for the 3 highest band-powers centered at 
$\ell_{\rm eff} \sim 660, 750, 850$.
This effect might be due to imperfect
foreground removal and/or the intrinsic estimator variance due to finite
volume and edge effects.
We estimated the dispersion in a set of WMAP simulations 
with Kp0 \& Kp2 sky cuts to be of the same order
as the measured differences in the $C_{\ell}$'s of the 
data. On the other hand, the cross-correlation amplitude
between the clean WMAP maps and the best fit foreground templates is at the 5\% and 10\% level of the WMAP $C_{\ell}$'s for the lowest and highest $\ell$'s, respectively.
We thus conclude that sample variance due to sky coverage 
can account for most of the observed difference in the $C_\ell$'s, 
while residual foreground contamination is always subdominant.
The low level of systematics in Kp2, and
the increased statistical errors due to the decreased sky fraction left by
Kp0, motivate us to adopt Kp2 
(as in H03) for the best estimate of the $C_{\ell}$'s.

\section{Error Estimation}
\label{sec:errors}

In order to estimate the covariance of our $C_\ell$'s, we generated MC  
simulations of the CMB sky and instrument noise for each of the 
8 DA's (Q1 through W4). We
used the {\sl running index} \lcdm model that best fits
a combination of WMAP, CBI \& ACBAR data (denoted {\sl WMAPext} in 
\cite{SpergelEtal2003}). 
Maps were convolved with the symmetric (non-Gaussian) beam
transfer function for each DA \citep{PageEtal2003}. 
As for the noise simulations, we downloaded 100 sky maps per DA 
from the LAMBDA website. 
These simulate 1 full year of flight instrument noise 
and they include all known radiometric effects 
\citep{HinshawEtal2003b,JarosikEtal2003}. 
Simulations were analyzed in exactly the same way as the data (see \S\ref{sec:ps}).
All in all, we have constrained the
errors from $1500$ measurements in MC simulations (combining cross spectra using 100 MC's for 
each of the 6 highest frequency DA's V1 through W4)
for the multipole ranges $\ell < 200$ and $\ell > 450$, 
and $2800$ measurements (combining cross-spectra using 100 MC's from each of the 8 DA's Q1 through W4) 
for the intermediate $\ell$-range $200 < \ell < 450$.

\begin{figure}[htb]
\figurenum{2}
\epsscale{1.}
\plotone{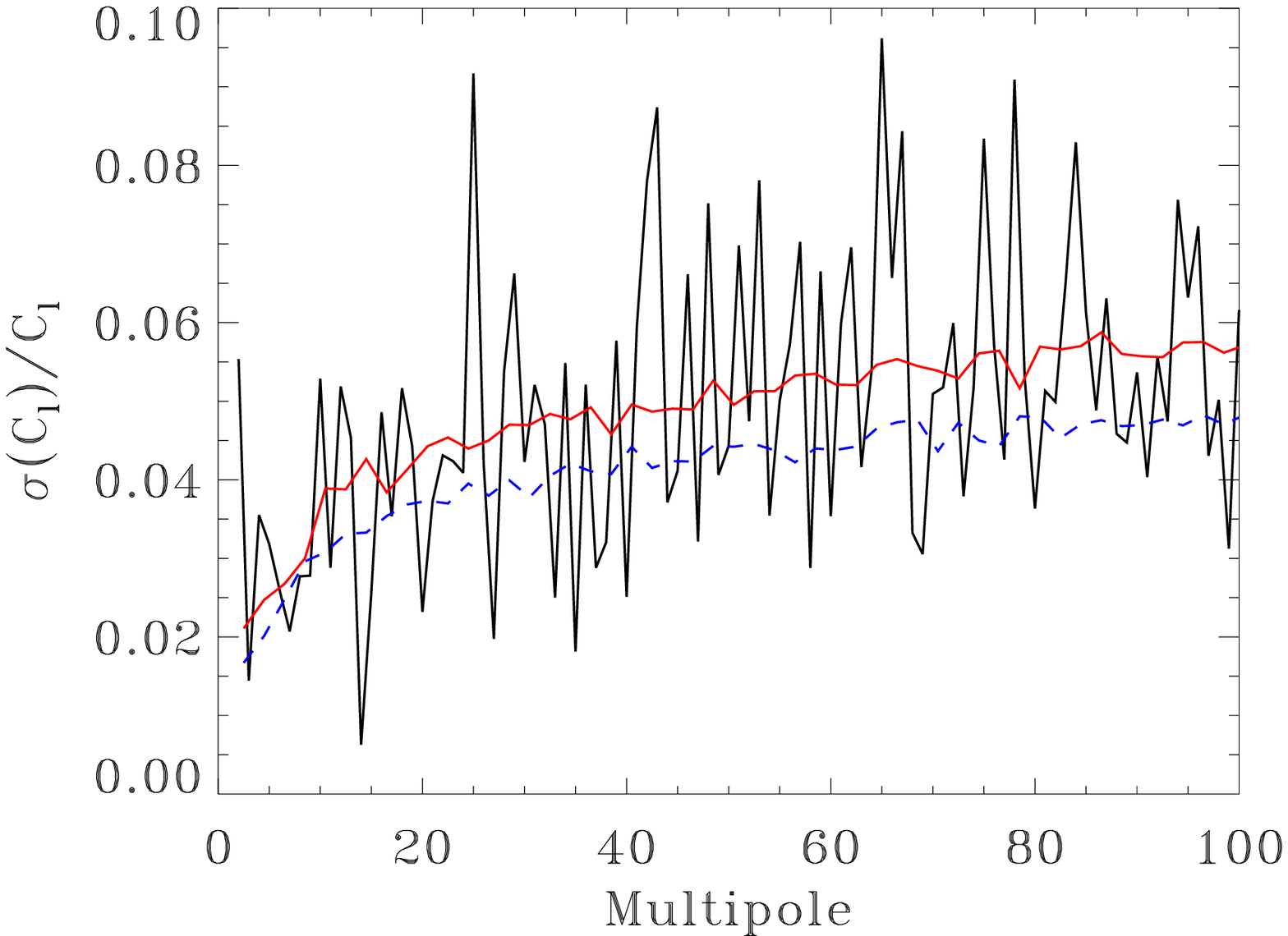}
\caption{Errorbars estimated from uncorrelated (dash line) and
correlated noise (solid smooth line) are shown. The spiky
solid line shows the variance of the $C_\ell$'s among the 
data channels, what should depend on noise only. The Monte Carlo
simulations with non-uniform noise appear to describe well the mean rms errors.}
\label{fig:lowlerr}
\end{figure} 

For multipoles $\ell<350$, errors in the WMAP power spectrum are dominated 
by cosmic or 
sample variance (see H03) and the noise only contributes 
at the few percent level.
Figure~\ref{fig:lowlerr} displays the noise contribution
to the relative errors at low multipoles, $\ell \simlt 100$.
Correlated noise simulation results are displayed
(smooth solid line) along with results from 
uncorrelated noise simulations (dashed line). 
The latter tends to underestimate errors by $\sim 1\%$.
Alternatively the noise level can be estimated from the data 
rms dispersion among the WMAP channels used (oscillating solid line). 
These results are in excellent agreement 
with H03 (cf. lower panel in their Figure~4).

At higher $\ell$'s  pixel noise and systematic effects 
(\eg beam and mode coupling, residual foregrounds)
increasingly dominate the errors.
MC methods assume detailed knowledge of all such
effects. To provide a model independent check of the errors,
we introduce a novel technique that allows estimating errors directly from the data:
the {\sl intra-bin variance} (IBV) method.
IBV estimates the variance of a given $C_{\ell}$ from
the rms dispersion in a bin $\rm B_{\ell}$ centered on $\ell$.
The bin-width ${\Delta \ell}$ is a 
matter of practical consideration, balancing variance and bias.
More precisely, our estimator for $\sigma(C_{\ell})$ reads
\begin{equation}
\sigma^2(C_{\ell}) = {1 \over {\Delta {\ell} -1}}  
\sum_{{\ell}^{\prime} \in {\rm B_{\ell}}}
({\Delta C_{{\ell}^{\prime}}}-{\langle{\Delta C_{\ell}\rangle}})^2
\label{eq:ibv}
\end{equation} 
where $\langle{\Delta C_{\ell}} \rangle = 1/\Delta {\ell} \sum_{{\ell}^{\prime} \in \rm B_{\ell}} \Delta C_{{\ell}^{\prime}},\,$
$\Delta C_{\ell} = {\bar C_{\ell}} - C^{\rm th}_{\ell}$, 
${\bar{C_{\ell}}}$ is the mean of the measured $C_{\ell}$'s over channels, 
and  $C^{\rm th}_{\ell}$ is 
our best guess for the data mean using a theoretical \lcdm model. 
The latter is subtracted to decrease the bias due to the
slope of the angular power spectrum. 
We used $C^{\rm th}_{\ell}$ from the WMAP best-fit {\sl running index} 
\lcdm model
\citep{SpergelEtal2003}, 
although this is not critical: no baseline subtraction only biases 
at a few percent level.
By construction, IBV should not be used to obtain errors with high
resolution but to
assess the overall level of errors in
a range of $\ell$'s, typically larger than $\Delta \ell$.

\begin{figure}[htb]
\figurenum{3}
\epsscale{1.}
\plotone{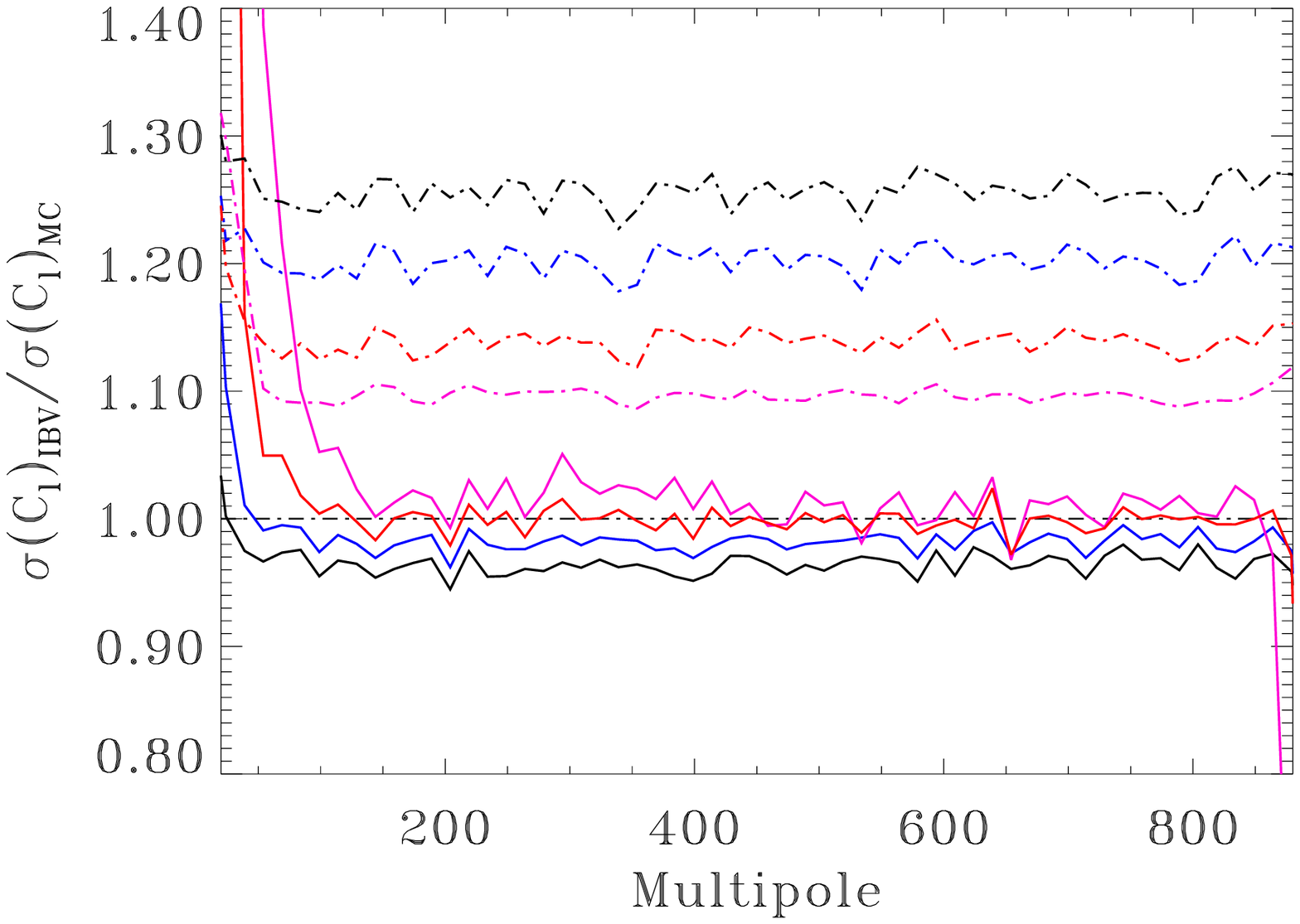}
\caption{Solid lines show the ratio of intra-bin variance (IBV) to Monte Carlo 
errors measured in WMAP simulations. At $\ell\simeq 300$
lines correspond to $\Delta\ell = 6,9,18,36$ (bottom up). The third
line ($\Delta\ell=18$) is unbiased for $\ell > 100$.
Upper dotted lines (in reverse order to solid lines) 
correspond to the relative rms error on the estimated IBV errors 
(\eg in the convention used, $1.15$ means $15\%$ error).}
\label{fig:ibcal}
\end{figure}

Figure~\ref{fig:ibcal} shows the ratio between the {\sl mean} 
IBV rms dispersion to the usual MC rms dispersion, 
both estimated from $\sim 3000$ WMAP simulations of CMB 
signal and correlated noise. Narrow bins yield
slightly biased (under-)estimates of the MC error at few percent level, possibly
due to small mode-to-mode couplings. 
IBV method with $\Delta \ell = 18$ yields 
unbiased estimates of the error for WMAP simulations 
at the 1\% level for $\ell \simgt 100$.
Doubling $\Delta \ell$ introduces a slight high bias and significant
edge effects for low $\ell$'s that could be caused by
the residual slope of the $C_{\ell}$'s. 
The unbiased bin-width, $\Delta \ell = 18$,
with $15\%$ variance
is our choice for the WMAP error estimation.

\begin{figure}[htb]
\figurenum{4}
\epsscale{1.}
\plotone{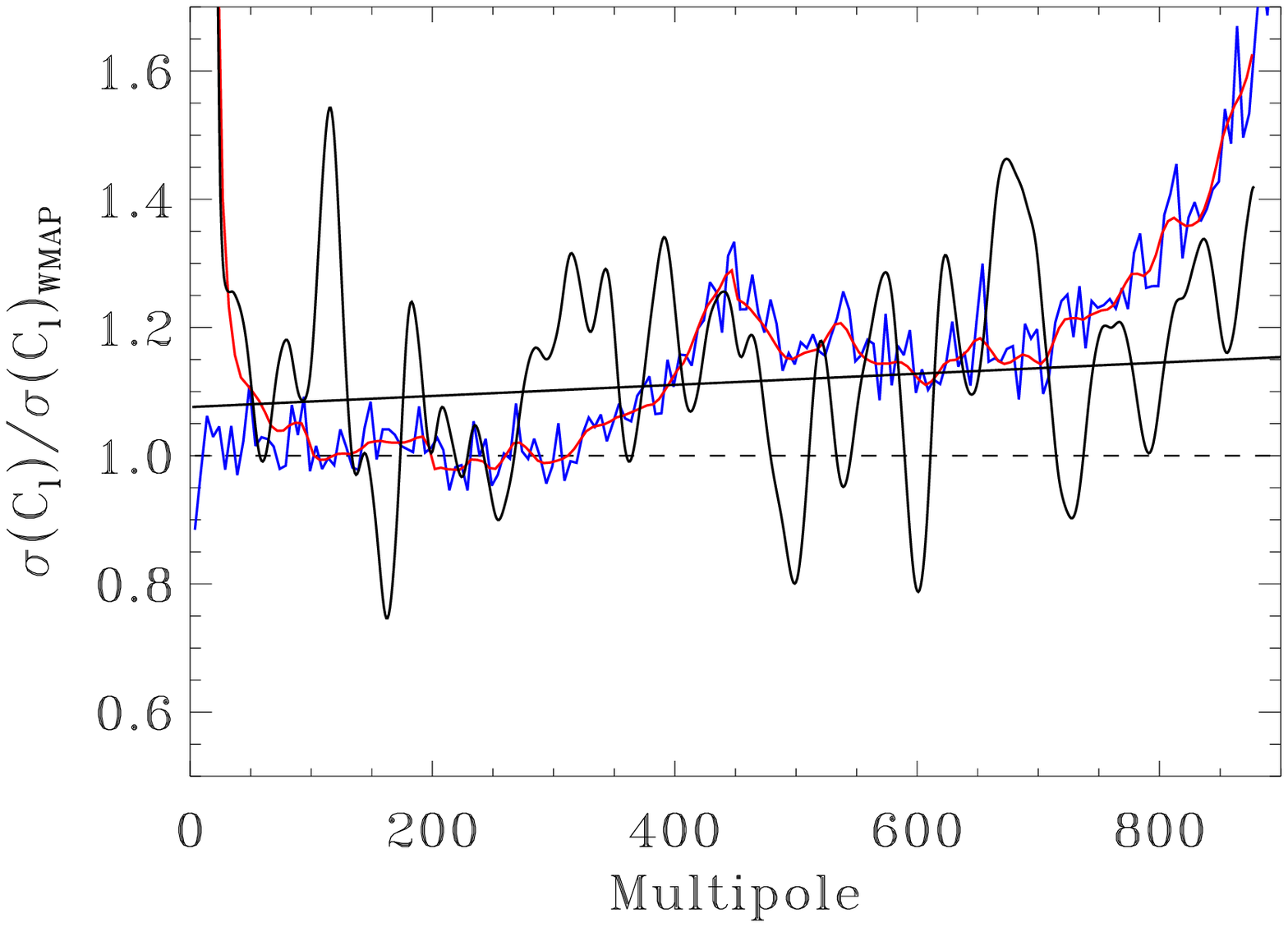}
\caption{Ratio of measured errors vs. published WMAP
errors \citep{VerdeEtal2003,HinshawEtal2003b,KogutEtal2003}. 
The noisy blue line and smooth red line through the latter
show the usual MC dispersion and the average IBV error respectively, 
for the same set of simulations. The large amplitude oscillating black line 
shows the IBV estimator applied to the WMAP
data. The IBV method detects 8-15\% larger errors than the
published ones. A conservative IBV error estimation
is depicted by the solid straight line obtained from
a linear least square fit.}
\label{fig:ibdata}
\end{figure} 

Figure~\ref{fig:ibdata} displays the WMAP data diagonal errorbars computed with the IBV 
method (spiky solid line) compared to the previously published 
diagonal errors (\cite{VerdeEtal2003},\cite{HinshawEtal2003b},\cite{KogutEtal2003}).  
The largest IBV errors appear to correlate well
with the outliers of the data $C_{\ell}$'s with
respect to the best-fit \lcdm model  (see Figure~3 in \cite{SpergelEtal2003}), suggesting
that our IBV estimator is closely related to a diagonal $\chi^2$ test.
It is clear that the mean overall error is higher than originally estimated (otherwise the IBV
curve would fluctuate around unity). The simplest and most conservative interpretation
of our results yields a monotonic error increase with respect to the 
WMAP team diagonal errors of the form,
$\sigma(C_{\ell})_{IBV}/\sigma(C_{\ell})_{WMAP} \simeq 1.08 + 8.5\cdot 10^{-5} ({\ell}-100)$
for $\ell>100$ (straight solid line in Figure~\ref{fig:ibdata}). This smooth
prediction results from a least squares minimization to the IBV curve
(large amplitude oscillating line in Figure~\ref{fig:ibdata}).
Note that for $\ell > 450$, the error excess is consistent with the
errors estimated from MC simulations with {\sl correlated} instrument noise 
(see noisy line in Figure~\ref{fig:ibdata} growing from left to right). 

In the range $100 < \ell < 450$ the mean error level is 
incompatible with both MC simulations that include correlated noise and the WMAP team published errors: 
given that there are approximately 16 independent $\Delta {\ell}$ bins, 
with an intrinsic 15 \% error each, and that the mean error excess is 9 \% in this $\ell$-range,
this amounts to a 2.4 $\sigma$ detection of the error excess.
We have checked that using the $C_{\ell}$'s measured by \cite{HinshawEtal2003a} yields 
comparable IBV errors in this multipole range.  
The excellent $\ell$-by-$\ell$ agreement between the SpICE and WMAP team's 
measurement of the $C_{\ell}$'s
(see Figure~\ref{fig:cls}) indicates that both estimators window functions are virtually identical 
in this regime, and thus the observed error excess points to systematics 
unaccounted for in the WMAP team analyses.
For $\ell > 450$ the interpretation is less clear
as there are hints that both window functions 
might be slightly different (see lower panel in Figure~\ref{fig:cls}).
Such differences might arise in the practical implementation of the estimators.
We also estimate a $\sim 5\%$ correlated noise contribution at $\ell<100$ 
(see Figure~\ref{fig:lowlerr}),  
that was neglected in previous likelihood analyses.
A more robust assessment of errors is provided below using the full $\chi^2$ test, 
where off-diagonal
terms are also taken into account following  Eq.(15) in \cite{VerdeEtal2003}.

\section{Discussion: Cosmological Parameters}

We investigate the implications of our measurements
using a Bayesian analysis of cosmological parameter estimation. 
We use CosmoMC\footnote{\rm http://cosmologist.info/cosmomc}, 
a Markov Chain Monte Carlo (MCMC) implementation \citep{LewisBridle2002}
based on CAMB\footnote{\rm http://camb.info} (\cite{LewisEtal1999}; 
see also CMBFAST\footnote{\rm http://cmbfast.org}, \cite{SeljakZaldarriaga1996}).
In order to allow direct comparison with \cite{SpergelEtal2003}, we  focus
on the simplest 6-parameter cosmological model consistent
with the WMAP temperature and cross-polarization data. 
Following \cite{VerdeEtal2003}, we assume a set of flat 
\lcdm models 
with radiation, baryons, cold dark matter and cosmological constant. Primordial
fluctuations are taken to be adiabatic and Gaussian with a 
power-law power spectrum.
We use the physical dark matter $\Omega_{cdm} h^2$ and baryon $\Omega_b h^2$ 
densities, the reionization optical depth $\tau$, the scalar spectral 
index $n_s$,
the normalized Hubble constant $h$, and the dark matter power spectrum 
normalization $\sigma_8$ \citep{KosowskyEtal2002}.
We estimate paramaters by combining 
4 independent chains with 30000 accepted points each, 
and use the 6 paramater covariance matrix as proposal density from precomputed
runs. This yields an excellent convergence-mixing 
Gelman \& Rubin statistic $R-1 \simlt 0.02$ for all cases studied.

\begin{deluxetable}{lll}
\tablecaption{Best Fit Parameters for Power Law $\Lambda$ CDM$^a$
\label{tab:bestfit}}
\tablewidth{0pt}
\tablehead{
\colhead{} & \colhead{SpICE $C_{\ell}$'s} & \colhead{WMAP $C_{\ell}$'s} \\
\colhead{} & \colhead{IBV Errors$^{b}$} & \colhead{Standard Errors$^{c}$}}
\startdata
$\tau$ &\ensuremath{0.145 \pm 0.067} &
\ensuremath{0.151 \pm 0.069} \\
$n_s$ &\ensuremath{0.99 \pm 0.04} &
\ensuremath{0.99 \pm 0.04} \\
$h$ &\ensuremath{0.67 \pm 0.05} &
\ensuremath{0.70 \pm 0.05} \\
$\Omega_b h^2$ &\ensuremath{0.0218 \pm 0.0014} &
\ensuremath{0.0234 \pm 0.0013} \\
$\Omega_{cdm} h^2$ &\ensuremath{0.122 \pm 0.018} &
\ensuremath{0.123 \pm 0.017} \\
$\sigma_8$ &\ensuremath{0.92 \pm 0.12} &
\ensuremath{0.92 \pm 0.11} \\
$\chi^2_{eff}/dof$ &1398.8/1342  &1428.7/1342  \\
\enddata
\tablenotetext{a}{WMAP Data Only. We impose a prior $\tau<0.3$. 
Table values are mean expectation values for the marginalized distribution
and errors are the $68 \%$ (symmetrized) confidence intervals.
}
\tablenotetext{b}{Parameters estimated with our MCMC's using
$C_{\ell}$'s measured with SpICE and IBV errors (see \S\ref{sec:errors}).}
\tablenotetext{c}{Same as b, but
using $C_{\ell}$'s from H03 and 
errors from \cite{VerdeEtal2003,HinshawEtal2003b,KogutEtal2003}.}
\end{deluxetable}

Table~1 summarizes our results. Imposing the prior $\tau < 0.3$, 
we find best fit values matching those of \cite{SpergelEtal2003}. In particular 
we obtain a $\chi^2/d.o.f. = 1.042$ (\ie it has a 14 \% probability) for the best-fit model 
(see first column in Table~1), 
which is a slightly better fit to the data than that of 
\cite{SpergelEtal2003}, $\chi^2/d.o.f. = 1.066$ (\ie 5 \% probability).
Our $h$ and $\tau$ are slightly lower but still consistent at the $1-\sigma$ level. 
This is more significant for our
estimates of the $C_{\ell}$'s and errors (see first column in Table~1).
In particular, our measurement $\Omega_b h^2 = 0.0218 \pm 0.0014$ agrees
with that from the latest BBN results $\Omega_b h^2 = 0.022 \pm 0.002$
\citep{CyburtEtal2003,VangioniEtal2003,CuocoEtal2003}.
We have checked that relaxing the $\tau$ prior yields larger values of
$\tau = 0.19 \pm 0.12$ \citep[cf.][]{TegmarkEtal2004}.
Our main results (see first column in Table~1) are 
in excellent agreement with the best-fit values from WMAP+SDSS \citep{TegmarkEtal2004},
and suggest a redshift of (abrupt) reionization $z_{re} = 16 \pm 5$ ($68\%$ CL).
Data products and additional plots from this work can be found at
{\rm http://www.ifa.hawaii.edu/cosmowave/wmap.html}

We thank an anonymous referee for insightful comments, 
Jun Pan for help and discussions,
Olivier Dore, Hans K. Eriksen, Eiichiro Komatsu for useful comments, 
and Antony Lewis for help with CosmoMC. 
We acknowledge extensive use of the 
Legacy Archive for Microwave Background
Data Analysis (LAMBDA). Support for LAMBDA is provided by the NASA
Office of Space Science.
Some of the results in this paper have been derived 
using HEALPix \citep{GorskiEtal1998}.
This research was supported by NASA through 
ATP NASA NAG5-12101 and AISR NAG5-11996, 
as well as by NSF grants AST02-06243 and ITR 1120201-128440.






\end{document}